\documentclass[aps,preprint,11pt,nofootinbib,a4paper,prd]{revtex4-1}
\usepackage{latexsym,bm,amsmath,amssymb,amsfonts}

\usepackage{graphicx,shortvrb,dblfloatfix}
\usepackage{epsfig}
\usepackage{footmisc}
\usepackage{setspace}
\usepackage{newlfont}
\usepackage{hyperref}
\usepackage{xcolor}
\usepackage{cool}
\usepackage{tikz,pgfplots}
\usetikzlibrary{positioning,shapes,shadows,arrows,svg.path}
\usepackage{booktabs}
\hypersetup{
    colorlinks,
    linkcolor={red!50!black},
    citecolor={blue!50!black},
    urlcolor={blue!80!black}
}

\usepackage{soul}

\AtBeginDocument{
\heavyrulewidth=.08em
\lightrulewidth=.05em
\cmidrulewidth=.03em
\belowrulesep=.65ex
\belowbottomsep=0pt
\aboverulesep=.4ex
\abovetopsep=0pt
\cmidrulesep=\doublerulesep
\cmidrulekern=.5em
\defaultaddspace=.5em
}

\begin{document}

\title{Stationary Lifshitz Black Hole of New Massive Gravity}

\author{{\"O}zg{\"u}r Sar{\i}o\u{g}lu}
\email{sarioglu@metu.edu.tr}
\affiliation{Department of Physics, Faculty of Arts and  Sciences,\\
              Middle East Technical University, 06800, Ankara, Turkey}

\date{\today}

\begin{abstract}
I present the stationary Lifshitz black hole solution of three-dimensional New Massive Gravity 
theory and study its elementary geometric and thermodynamical properties.
\end{abstract}

\maketitle

\section{Introduction}\label{intro}
As well known, the celebrated AdS/CFT correspondence has also been utilized in non-relativistic
condensed matter systems \cite{Son:2008ye,Hartnoll:2009sz} (and the references therein).
In this regard, the anisotropic scaling symmetry (also called the \emph{Lifshitz symmetry})
\[ t \mapsto \lambda^{z} \, t \,, \quad \rho \mapsto \frac{\rho}{\lambda} \,, \quad 
\vec{x} \mapsto \lambda \, \vec{x} \,, \]
where $z>1$ is called the \emph{dynamical exponent}, has been successfully imposed on
boundary field theories with a corresponding bulk described by the static Lifshitz spacetime 
metric \cite{Kachru:2008yh} 
\begin{equation}
ds^2 = - \frac{\rho^{2z}}{\ell^{2z}} dt^2 + \frac{\ell^2}{\rho^2} d\rho^2 
+ \frac{\rho^2}{\ell^2} \, d\vec{x}^2 \,. \label{lifgen}
\end{equation}
On the gravity side, one either needs (various types of) matter couplings 
\cite{Kachru:2008yh,Taylor:2008tg,Danielsson:2009gi} 
and/or higher curvature models \cite{AyonBeato:2009nh,Cai:2009ac} to support the Lifshitz 
spacetime (\ref{lifgen}) and/or analytic or numerical Lifshitz black hole and black brane solutions. 
In this regard, black hole solutions are special since they describe the finite temperature behavior
of the dual non-relativistic field theories. However, there are only a few \emph{exact} static 
and, even less number of, stationary Lifshitz black holes known 
\cite{AyonBeato:2009nh,Cai:2009ac}. 

This work provides an important addition to this modest list of exact Lifshitz black holes: I
present the stationary Lifshitz black hole of three-dimensional New Massive Gravity (NMG) 
theory, and study its basic geometric and thermodynamical properties. It is worth emphasizing 
that this solution can be used as a test case for investigating discrepancies between various 
methods for calculating gravitational charges of spacetimes with non-standard (in particular, 
non-AdS, and in general, anisotropic) asymptotics relevant for generalized (especially, 
non-relativistic) holography.

Briefly stated, NMG, the gravitational model of interest in this work, was originally introduced
\cite{Bergshoeff:2009hq} as a \emph{parity-preserving} and \emph{unitary} solution to the 
problem of consistently extending the Fierz-Pauli field theory for a massive spin-2 particle 
to include interactions. To this end, the source-free NMG action was obtained by adding 
a specially-chosen quadratic term to the cosmological Einstein-Hilbert piece and reads 
\cite{Bergshoeff:2009hq}
\begin{equation}
I_\mathrm{NMG} = \int d^3 x \, \sqrt{-g} \, \Big( R - 2 \Lambda_{0} + 
\frac{1}{m^2} \big( S^{\mu\nu} S_{\mu\nu} - S^2 \big) \Big) \,. \label{nmg}
\end{equation}
Here $\Lambda_{0}$ is the cosmological constant (with dimensions 1/Length$^2$), $m$ is 
a mass parameter (with dimensions 1/Length), and the Schouten tensor $S_{\mu\nu}$ and 
its trace $S$ are given by
\begin{equation}
S_{\mu\nu} \equiv R_{\mu\nu} - \frac{1}{4} R \, g_{\mu\nu}  \,, \qquad 
S \equiv g^{\mu\nu} S_{\mu\nu} = \frac{R}{4} \,.
\label{schten}
\end{equation}

The organization of the paper is as follows: In section \ref{spaces}, I show how the
stationary Lifshitz black hole can be obtained from the static Lifshitz black hole by
a simple boost, and discuss its basic geometric properties. Section \ref{thermo} is devoted
to the calculation of the thermodynamical quantities of the stationary Lifshitz black hole, and on 
how the first law of black hole thermodynamics can be utilized to conjecture the energy and the
angular momentum of this black hole. I then finish up with a discussion of possible future work.
I give the technical details on the properties of the cubic polynomial that 
is essential for the derivation of the stationary Lifshitz black hole in appendix \ref{appa},
and examine the geodesics of the stationary Lifshitz black hole and compare them to the 
geodesics of the static one in appendix \ref{appb}. 

\section{Lifshitz spacetimes and Lifshitz black holes of NMG}\label{spaces}
The field equation that follows from the variation of the action (\ref{nmg}) is
\begin{equation}
R_{\mu\nu} - \frac{1}{2} R \, g_{\mu\nu} + \Lambda_{0} \, g_{\mu\nu} + \frac{1}{m^2} K_{\mu\nu} = 0 \,, 
\label{NMG}
\end{equation} 
where 
\( K_{\mu\nu} \equiv \square S_{\mu\nu} - \nabla_{\mu} \nabla_{\nu} S 
+ S S_{\mu\nu} - 4 S_{\mu\rho} S_{\nu}\,^{\rho} + \tfrac{1}{2} g_{\mu\nu} 
\big( 3 S^{\rho\sigma} S_{\rho\sigma} - S^2 \big). \) It was shown in \cite{AyonBeato:2009nh} that, 
for the special choice 
\begin{equation}
\Lambda_{0} = - \frac{13}{2 \ell^2} \,, \qquad m^2 = \frac{1}{2 \ell^2} \label{para}
\end{equation}
of the parameters, the \emph{static Lifshitz black hole}
\begin{equation}
ds^2 = - \frac{\rho^6}{\ell^6} \left( 1 - \frac{M \, \ell^2}{\rho^2} \right) dt^2 
+ \frac{d\rho^2}{\big( \frac{\rho^2}{\ell^2} -M \big)} + \rho^2 \, d\theta^2 
\label{stamet}
\end{equation}
is a solution of the NMG field equations (\ref{NMG}). Note that when the parameter $M$ is
set to zero in (\ref{stamet}), one is led to the \emph{static Lifshitz spacetime} 
(with dynamical exponent $z=3$)
\cite{Kachru:2008yh}
\begin{equation}
ds^2 = - \frac{\rho^6}{\ell^6} dt^2 + \frac{\ell^2}{\rho^2} d\rho^2 + \rho^2 \, d\theta^2 \,. \label{lif}
\end{equation}

Now let me rewrite (\ref{stamet}) by making the coordinate transformation $\rho^2 = x$ as
\begin{equation}
ds^2 = - \frac{x^3}{\ell^6} \left( 1 - \frac{M \, \ell^2}{x} \right) dt^2 
+ \frac{dx^2}{4 \, x \, \big( \frac{x}{\ell^2} -M \big)} + x \, d\theta^2 \,,
\label{stametinx}
\end{equation}
and simply boost (\ref{stametinx}) via
\begin{equation} 
\left[ 
\begin{array}{c}
dt \\
d\theta
\end{array}
\right] \to
\frac{1}{\sqrt{1-\omega^2}}\left[
\begin{array}{lr}
  1 & -\omega \ell \\
  -\omega/\ell & 1 
\end{array}
\right]
\left[ 
\begin{array}{c}
d t \\
d \theta
\end{array}
\right] \,, \label{boost}
\end{equation}
where the ``rotation parameter" $\omega$ is a real constant with \( |\omega| < 1 \),
to arrive at the stationary metric 
\begin{eqnarray}
ds^2 & = & - \frac{dt^2}{(1-\omega^2)} 
\left( \frac{x^3}{\ell^6} - \frac{M \, x^2}{\ell^4} - \frac{\omega^2 \, x}{\ell^2} \right) 
+ \frac{2 \, \omega \, \ell \, dt \, d\theta}{(1-\omega^2)}  
\left( \frac{x^3}{\ell^6} - \frac{M \, x^2}{\ell^4} - \frac{x}{\ell^2} \right) \nonumber \\
& & + \frac{\ell^2 \, d\theta^2}{(1-\omega^2)} 
\left( \frac{x}{\ell^2} - \frac{\omega^2 \, x^3}{\ell^6} + \frac{M \, \omega^2 \, x^2}{\ell^4} \right)
+ \frac{dx^2}{4 \, x \, \big( \frac{x}{\ell^2} -M \big)} \,. \label{altmet}
\end{eqnarray}
Let me identify the coefficient of the \( d\theta^2 \) term in (\ref{altmet}) by introducing
the coordinate transformation $x=x(r)$ that is described by the cubic polynomial\footnote{The 
properties of the cubic polynomial (\ref{cubic}) are studied in detail in appendix \ref{appa}. 
The existence of at least one real root $x(r)$ is guaranteed of course.}
\begin{equation}
x + \frac{M \, \omega^2}{\ell^2} x^2 - \frac{\omega^2}{\ell^4} x^3 - (1-\omega^2) r^2 = 0 \,,
\label{cubic}
\end{equation}
such that (\ref{altmet}) can be written as
\begin{equation}
ds^2 = - \frac{dt^2}{\omega^2 \, \ell^2} \big( (1+ \omega^2) \, x(r) - r^2\big) 
- \frac{2 \, dt \, d\theta}{\omega \, \ell}  \big( r^2 - x(r) \big) + r^2 \, d\theta^2 
+ \frac{(x^{\prime})^2 \, dr^2}{4 \, x(r) \, \big( \frac{x(r)}{\ell^2} -M \big)} \,. \label{met}
\end{equation}
Here prime denotes differentiation with respect to the coordinate $r$. Since the metric 
(\ref{met}) and the polynomial (\ref{cubic}) are both left invariant under \( r \mapsto -r \) 
(and independently under \( \omega \mapsto -\omega \)), one can think of the variables 
$(r,\theta)$ as the polar coordinates on the Euclidean plane with the ranges \( r \geq 0 \) 
and \( \theta \in [0, 2 \pi) \), and assume $0 \leq \omega<1$ without loss of generality.
Here the temporal coordinate $t$ takes any real value \( t \in \mathbb{R} \), and the metric 
is circularly-symmetric with Killing vectors \( (\partial/\partial t)^{\mu} \) and 
\( (\partial/\partial \theta)^{\mu} \).

Note that even though the static Lifshitz black hole (\ref{stamet}) and, of course, the static
Lifshitz spacetime (\ref{lif}) enjoy the Lifshitz scaling symmetry 
\( (t \mapsto \lambda^3 \, t , \rho \mapsto \rho/ \lambda , \theta \mapsto \lambda \, \theta) \)
provided \( M \mapsto M/\lambda^2 \) as well \cite{AyonBeato:2009nh}, this is no longer so 
for the stationary metric (\ref{altmet}) (with the understanding that \( x \mapsto x/\lambda^2 \)).

As a side remark, one can also keep $\omega$ ``on", i.e. \( \omega \neq 0 \), but switch-off $M$
in (\ref{altmet}), that is boost the static Lifshitz spacetime (\ref{lif}) (written in terms of the variable
$x$) by (\ref{boost}), to arrive at what I will call as \emph{stationary Lifshitz spacetime}\footnote{Here
the token `stationary Lifshitz spacetime' is an obvious misnomer since switching-off
$M$ does not help in restoring the Lifshitz scaling symmetry lost with the turning-on of $\omega$.
Please refer to the last sentence of the next paragraph for the rationale behind this choice.}. 
Going backwards, this results merely in setting $M=0$ in (\ref{met}) and (\ref{cubic}), of course.

The stationary metric (\ref{met}) (or (\ref{altmet})) has a curvature singularity at $r=0$ (or $x=0$), 
which can also be seen from the curvature invariants
\[ R = - \frac{26}{\ell^2} + \frac{8 M}{x(r)} \,, \qquad 
R_{\mu\nu} R^{\mu\nu} = 
4 \left( \frac{65}{\ell^4} - \frac{38 M}{\ell^2 x(r)} + \frac{6 M^2}{x^2(r)} \right) \,. \] 
As shown explicitly in appendix \ref{appa}, the cubic polynomial (\ref{cubic}) is guaranteed 
to have at least one real root, so the function $x(r)$ in (\ref{met}) is indeed well-defined. 
The $rr$-component of (\ref{met}) diverges when \( x(r) = \ell^2 M \) and for $M>0$ this leads 
to the coordinate singularity at \( r_{+} \equiv \ell \sqrt{M/(1-\omega^2)} > 0 \), describing the 
event horizon. Thus I call the metric (\ref{met}) (or equivalently (\ref{altmet})\footnote{For 
the interpretation of (\ref{altmet}) still as a black hole, one must implicitly assume that 
the coordinate $x$ in (\ref{altmet}) is in the range \( 0 < x < x_{+} \) (\ref{xpmdef}). One must not
take (\ref{altmet}) on its own and assume wrongly that \( x \in \mathbb{R} \) here.}) as the 
\emph{stationary Lifshitz black hole} of NMG, even though it is \emph{neither} left invariant 
under the `proper' Lifshitz scalings as pointed out earlier \emph{nor} 
asymptotically Lifshitz as one formally takes $x$ (or $r$) $\to \infty$\footnote{See the 
previous footnote on the range of the variable $x$.}, since it derives from 
the static Lifshitz black hole (\ref{stamet}).

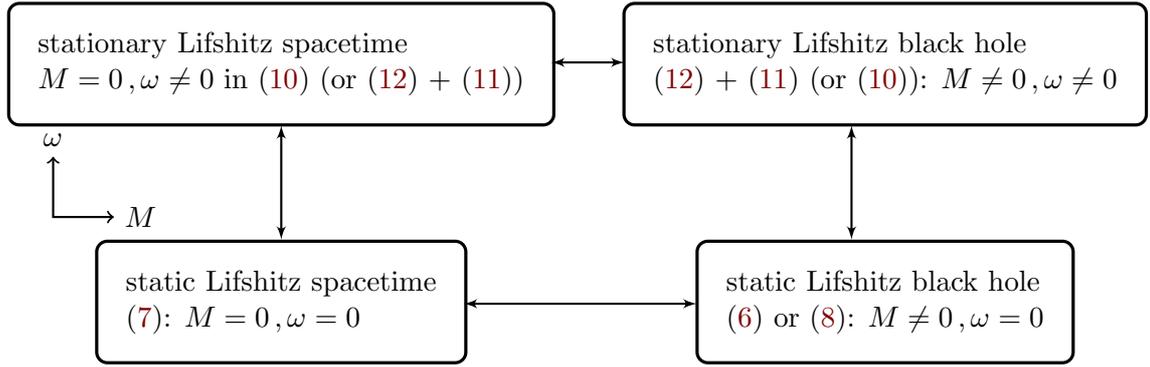
\begin{figure}[t]
\centering
\begin{tikzpicture}[node distance=1cm, auto]  
\tikzset{
    mynode/.style={rectangle,rounded corners,draw=black,top color=white,bottom color=white!50,
    very thick, inner sep=1em, minimum size=3em, text centered},
    myarrow/.style={<->, >=latex', shorten >=1pt, thick}, 
}  
\node[mynode,align=left,below] (static) {static Lifshitz spacetime \\ (\ref{lif}): \( M=0 \,, \omega=0 \)};
\node[mynode,align=left,below,above=1.5cm of static] (stationary) {stationary Lifshitz spacetime \\ 
\( M=0 \,, \omega \neq 0 \) in (\ref{altmet}) (or (\ref{met}) + (\ref{cubic}))};
\node[mynode,align=left,below,right=3cm of static] (statBH) {static Lifshitz black hole \\ 
(\ref{stamet}) or (\ref{stametinx}): \( M \neq 0 \,, \omega=0 \)};
\node[mynode,align=left,below,above=1.5cm of statBH] (rotBH) {stationary Lifshitz black hole \\ 
(\ref{met}) + (\ref{cubic}) (or (\ref{altmet})): \( M \neq 0 \,, \omega \neq 0 \)};

\draw[myarrow] (2.4,-0.85) -- (5.5,-0.85);
\draw[myarrow] (3.55,2.35) -- (4.55,2.35);

\draw[myarrow] (0,0) -- (0,1.55);
\draw[myarrow] (7.5,0) -- (7.5,1.55);

\draw [<->,thick] (-3,1.1) node (waxis) [above] {$\omega$} -- (-3,0.3) -- (-2.2,0.3) node (Maxis) [right] {$M$};

\end{tikzpicture}
\medskip
\caption{The static and stationary Lifshitz spacetimes and Lifshitz black holes of NMG} 
\label{fig}
\end{figure}

For the sake of convenience, I summarize the four metrics described in this section, and the relations 
between them, in Fig. \ref{fig}, and present the stationary Lifshitz black hole explicitly using the 
$\rho$-coordinate as well:
\begin{eqnarray}
ds^2 & = & - \frac{dt^2}{(1-\omega^2)} 
\left( \frac{\rho^6}{\ell^6} - \frac{M \, \rho^4}{\ell^4} - \frac{\omega^2 \, \rho^2}{\ell^2} \right) 
+ \frac{2 \, \omega \, \ell \, dt \, d\theta}{(1-\omega^2)}  
\left( \frac{\rho^6}{\ell^6} - \frac{M \, \rho^4}{\ell^4} - \frac{\rho^2}{\ell^2} \right) \nonumber \\
& & + \frac{\ell^2 \, d\theta^2}{(1-\omega^2)} 
\left( \frac{\rho^2}{\ell^2} - \frac{\omega^2 \, \rho^6}{\ell^6} + \frac{M \, \omega^2 \, \rho^4}{\ell^4} \right)
+ \frac{d\rho^2}{\big( \frac{\rho^2}{\ell^2} -M \big)} \,. \label{altmetinrho}
\end{eqnarray}
Finally, for `geometers at heart' I briefly discuss the geodesics of the stationary 
Lifshitz black hole (\ref{altmetinrho}) and compare these with the geodesics of the static 
Lifshitz black hole (\ref{stamet}) in appendix \ref{appb}.

\section{Thermodynamics of the stationary Lifshitz black hole}\label{thermo}
I now turn to the question/challenge of examining the thermodynamics of the stationary Lifshitz 
black hole. For that purpose, it is useful to review the analogous properties of the static Lifshitz 
black hole first.

It was in \cite{Hohm:2010jc} that the energy of the static Lifshitz black hole was calculated first. 
The authors of \cite{Hohm:2010jc} employed the so-called boundary stress tensor method, but 
did so with a non-trivial caveat: The counterterm they obtained was not uniquely determined; 
one could put, at best, two physical conditions to determine the three free parameters that needed 
to be fixed. The authors of \cite{Hohm:2010jc} had to resort to the validity of the first law of 
thermodynamics, \( dE = T dS \), to get over this ambiguity. In hindsight, it is easy to see that one 
could in fact do away with the boundary stress tensor method altogether, calculate the temperature 
$T$ and the entropy $S$ through standard methods (e.g. using the Wald entropy \cite{Iyer:1994ys}) 
and arrive at the energy of the static Lifshitz black hole with relative ease. 

As an alternative, the authors of \cite{Myung:2009up} have instead performed a dimensional
reduction (by exploiting the circular symmetry of the static Lifshitz black hole (\ref{stamet})) of the 
NMG theory to arrive at a complicated two-dimensional dilaton gravity, studied the properties
of the analogous black hole obtained so in two dimensions, and showed that the thermodynamics 
of the original black hole in three dimensions could be consistently derived from there. The upshot 
of both of these calculations is that (in the units that I adopt in this work) the relevant 
thermodynamic quantities of the static Lifshitz black hole read
\begin{equation}
T = \frac{M^{3/2}}{2 \pi \ell} \,, \quad S = \frac{2 \pi \ell \sqrt{M}}{G} \,, \quad
E = \frac{M^2}{4 G} \,, \label{thermolif}
\end{equation}
where $G$ denotes the three-dimensional Newton's constant, and indeed the first law of 
thermodynamics \( dE = T dS \) holds.

Recognizing the need for a more direct approach for the computation of conserved charges
(such as energy and angular momentum) of spacetimes that do not asymptote to spaces of
maximal symmetry (such as Minkowski or AdS spaces) but instead to exotic ones such as 
Lifshitz spaces, it was in \cite{Devecioglu:2010sf} that an extension of the conserved Killing
charge definition of the ADT procedure \cite{Deser:2002jk} was given and developed for quadratic
curvature gravity models in generic dimensions. There it was also shown that this extension is
background gauge invariant and reduces to the one in \cite{Deser:2002jk} when the background
is a space of constant curvature. To cut a long story short, application of this hands-on approach 
to the static Lifshitz black hole \cite{Devecioglu:2010sf,Devecioglu:2011yi} led to the energy
\begin{equation}
E = \frac{7 M^2}{8 G} \,, \label{wrongE}
\end{equation}
which is clearly different from (\ref{thermolif}) and not in accord with the first law of 
thermodynamics. This is quite discouraging, to say the least, but as discussed in 
\cite{Devecioglu:2011yi}, may stem from a number of reasons. Instead of going over that discussion
once more, let me point out to the most obvious one here: Any number of hypotheses,
especially ``the assumption that the deviations vanish sufficiently fast as one asymptotically 
approaches to the boundary of spacetime described by the background" and/or ``the applicability of
the Stokes' theorem on the relevant hypersurfaces and/or boundaries", which were crucial in 
the derivation of the extended definition of conserved charges \cite{Devecioglu:2010sf} at the first
place, may be violated by exotic spacetimes such as Lifshitz black holes.

Despite the disappointment in the discrepancy between (\ref{thermolif}) and (\ref{wrongE}), 
the extended definition of the Killing charge was quite successful in directly determining the 
energy of the warped AdS black hole solution of NMG \cite{Clement:2009gq}, but unfortunately 
led to a slightly different expression for the angular momentum (see \cite{Devecioglu:2011yi} for 
details). One of the main motivations of the present work has been to find a new concrete and
\emph{stationary} example where the extended Killing charge definition \cite{Devecioglu:2010sf} 
could be applied to, apart from the warped AdS black hole. 

One may question why I am ``insisting on" using the theoretical approach advanced in
\cite{Devecioglu:2010sf} when it has already failed in a number of cases as explained above.
The plain reason is that it gives the ``correct" structural form of the conserved quantities and
falls flat only at the numerical factors in front of the charges. In what follows, I want to extract,
at least, the forms of the energy $E$ and the angular momentum $L$ for the stationary Lifshitz 
black hole, see the effect of turning on the parameter $\omega$ and speculate, if necessary, on 
the ``correct" $E$ and $L$ from there on.

After this informative digression, let me turn back to the problem of studying the thermodynamics
of the stationary Lifshitz black hole now. Using the definition for the angular velocity of the horizon
$\Omega_{H}$, the surface gravity $\varkappa$, thus the temperature $T$, and Wald entropy $S$
\cite{Iyer:1994ys}, I find the following 
\begin{equation}
\Omega_{H} = \frac{\omega}{\ell} \,, \quad 
T = \frac{\varkappa}{2 \pi} = \frac{M^{3/2}}{2 \pi \ell} \sqrt{1-\omega^2} \,, \quad
S = \frac{2 \pi \ell \sqrt{M}}{G \sqrt{1-\omega^2}} \label{OmTS}
\end{equation}
for elementary thermodynamical quantities. Here, I have suitably adapted the general discussion
given in \cite{Devecioglu:2011yi} to the conventions used throughout, i.e. I have set
\[ \kappa = 16 \pi G \,, \quad \Omega_{1} = 2 \pi \,, \quad 
\beta = \frac{1}{m^2 \kappa} = \frac{2 \ell^2}{\kappa} \,, \quad
\alpha = - \frac{3}{8} \beta \,, \quad \gamma = 0 \]
in \cite{Devecioglu:2011yi}. As for the calculation of the energy $E$ and the angular momentum $L$,
I again refer the reader to the detailed discussion given in Section IV of \cite{Devecioglu:2011yi}
(especially to the parts on the warped AdS black hole solution of NMG). A naive calculation using 
the static Lifshitz spacetime as background immediately leads to divergent $E$ and $L$. However, 
a close scrutiny makes it apparent that ``the fall-off conditions that need to be satisfied by the
deviations" (as alluded to earlier) are severely violated in this case. 
Keeping this observation in mind, it turns out that the most reasonable thing to do is to work with 
the stationary Lifshitz spacetime as background\footnote{One may naively call for the background
itself, i.e. the stationary Lifshitz spacetime in this case, to satisfy the very same fall-off conditions
as the deviations, but this is against the whole essence of this procedure. In a sense,
this background choice is the most natural one that ``renormalizes" the divergences encountered.}. 
Sparing the reader form the gory details of 
rather long calculations, I have summarized the outcome of the energy and the angular 
momentum calculations of the stationary Lifshitz black hole with respect to one of three 
physically sensible background choices in Table \ref{table}\footnote{I can briefly explain the 
steps taken though: One first calculates the ``extended Killing charge density" at finite $r$ 
for which $r>r_{+}$, the outcome of which is a rather long expression, better not displayed 
here, but an even, rational function of finite $r$. Then I assumed $r \gg 1$, which is plausible,
and ignoring the lesser powers of $r$ both in the numerator and the denominator led me to 
the results in Table \ref{table}. The relevant ones also reduced to what was earlier found in 
the $\omega \to 0$ limit.}. 
\begin{table*}[t]
\caption{\label{table} The energy and the angular momentum of the stationary Lifshitz black hole} 
\begin{ruledtabular}
\begin{tabular}{lll}
background & Energy $E$ & Angular Momentum $L$ \\ 
\midrule
static Lifshitz spacetime & divergent & divergent \\ 
stationary Lifshitz spacetime & \( \frac{M^2 (7 + 11 \omega^2)}{8 G (1-\omega^2)} \) & 
\( \frac{18 M^2 \ell \omega}{8 G (1-\omega^2)} \) \\
static Lifshitz black hole & divergent & \( \frac{4 M^2 \ell \omega}{8 G (1-\omega^2)} \) \\
\end{tabular} 
\end{ruledtabular} 
\end{table*}

Clearly one cannot have the first law of thermodynamics (in the form 
\( dE = T \, dS + \Omega_{H} \, dL \)) hold even by using the only reasonable pair
\begin{equation}
E = \frac{M^2 (7 + 11 \omega^2)}{8 G (1-\omega^2)} \,, \quad
L = \frac{18 M^2 \ell \omega}{8 G (1-\omega^2)} \label{wrongEL}
\end{equation}
here. Instead of giving up, let me do the following: Demand that 
i) \( dE = T \, dS + \Omega_{H} \, dL \) holds with (\ref{OmTS}), and that 
ii) all respective quantities approach their counterparts for the static Lifshitz black hole
when one takes \( \omega \to 0 \). Keeping the general features of both $E$ and $L$ intact, the
cheapest way to do so is by tweaking the coefficients in (\ref{wrongEL}), i.e. to take
\begin{equation}
E = \frac{M^2 (2 + a \, \omega^2)}{8 G (1-\omega^2)} \,, \quad
L = \frac{b \, M^2 \ell \omega}{8 G (1-\omega^2)} \,, \label{ELans}
\end{equation}
and later to determine the coefficients $(a,b)$ using the first of law of thermodynamics.
Doing so, one finds that there is only a unique nontrivial pair: $(a,b)=(6,8)$. For what it is
worth, I thus conjecture that the ``correct" energy and the ``correct" angular momentum 
of the stationary Lifshitz black hole is given by
\begin{equation}
E = \frac{M^2 (1 + 3 \, \omega^2)}{4 G (1-\omega^2)} \,, \quad
L = \frac{M^2 \ell \omega}{G (1-\omega^2)} \,. \label{ELcon}
\end{equation}

\section{Discussion}\label{disc}
In this note I have presented the stationary Lifshitz black hole of NMG, 
and studied its elementary geometric and thermodynamical properties. Even though the charges
calculated using the extended conserved Killing charge definition \cite{Devecioglu:2010sf} were 
not in accord with the first law of thermodynamics, assuming the validity of the first law 
(and of course taking the Wald entropy for granted) I predicted the energy and the angular 
momentum of the stationary Lifshitz black hole. 

It is a separate but, of course, a legitimate question to understand the ``physical meanings"
of the conserved quantities, since the asymptotic behavior, if any, of the metric (\ref{met}) is far 
from clear. (\ref{met}) must certainly be studied further, perhaps using numerical methods as 
well to alleviate the difficulties arising from the cubic polynomial (\ref{cubic}). A more detailed 
examination of the geodesics, initiated in appendix \ref{appb} here, would undoubtedly be of 
help in this endeavor.

The stationary Lifshitz black hole should allow for getting rid of the ambiguity encountered in
uniquely determining the three free parameters in the counterterm \cite{Hohm:2010jc} that 
emerged when trying to use the boundary stress tensor method. It is also worth trying to 
generalize the dimensional reduction developed in \cite{Myung:2009up} and to work out
the conserved charges of the stationary Lifshitz black hole from that side. One immediate
calculation that should be worth the effort is to compute the energy and the angular momentum
by using the ``quasilocal generalization" \cite{Kim:2013zha} of the conserved Killing charges 
method employed here. 

As stated in the text, one obvious source of error causing the theoretical procedure developed
in \cite{Devecioglu:2010sf} to fail is the violation of the fall-off conditions demanded from the 
deviations. Pending a computation, e.g. via the method of \cite{Kim:2013zha}, to check the
conjecture advanced in this work, one may contemplate devising a convenient cut-off 
mechanism and scrutinizing the effects, if any, of the boundary terms that are thrown away in the
derivation of the field equations to the application of the procedure given in \cite{Devecioglu:2010sf}.
It certainly is worth the effort to understand the cause of the inconsistency between the
conserved charge computation and the first law of black hole thermodynamics, and to find a 
solid solution to fix this problem. 

To recapitulate, this work has mainly focused on the presentation of the stationary Lifshitz 
black hole and understanding the energy and the angular momentum through a rather 
conventional manner. It must also be worth studying other geometric features as well as
physical properties and their consequences in the context of condensed matter physics via
the AdS/CFT correspondence and NMG holography \cite{Sinha:2010ai}.

\begin{acknowledgments}
I would like to thank Deniz Olgu Devecio\u{g}lu for a critical reading of the manuscript, and to
Gaston Giribet for a useful correspondence after the first draft of this manuscript appeared in the arXiv.
\end{acknowledgments}

\appendix
\section{\label{appa}} 
Here I want to examine the cubic polynomial (\ref{cubic}) in more detail. In what follows, I will take $0<\omega<1$ and $0<M$ to simplify the discussion on the black-hole interpretation of (\ref{met}).
Let me start by writing (\ref{cubic}) in the canonical form
\[ x^3 + a_2 x^2 + a_1 x + a_0 = 0 \,, \]
where I have defined the coefficients
\[ a_2 \equiv - M \ell^2 < 0 \,, \quad a_1 \equiv - \frac{\ell^4}{\omega^2} < 0 \,, \quad 
a_0 \equiv \frac{\ell^4}{\omega^2} (1-\omega^2) r^2 > 0 \,. \]
Using these, let me also introduce \cite{abrasteg}
\begin{eqnarray*}
Q & \equiv & \frac{a_1}{3} - \frac{a_2^2}{9} = - \frac{\ell^4}{9 \omega^2} (3 + M^2 \omega^2) < 0 
\,, \\
P & \equiv & \frac{a_2 a_1}{6} - \frac{a_0}{2} - \frac{a_2^3}{27} = 
\frac{\ell^4}{2 \omega^2} \Big( \frac{M \ell^2}{27} (9 + 2 M^2 \omega^2) - (1-\omega^2) r^2 \Big) \,, \\
\Delta & \equiv & Q^3 + P^2 = \frac{\ell^8}{108 \omega^6} \Big( 27 r^4 \omega^2 (1-\omega^2)^2 
- \ell^4 (4 + M^2 \omega^2) - 2 M r^2 \ell^2 \omega^2 (1-\omega^2) (9 + 2 M^2 \omega^2) \Big) \,. \\
\Xi & \equiv & \sqrt[3]{P+\sqrt{\Delta}} \,, \qquad \Upsilon \equiv \sqrt[3]{P-\sqrt{\Delta}} \,.
\end{eqnarray*}
Finally, the \emph{formal roots} of the polynomial (\ref{cubic}) are given by
\begin{eqnarray}
x_1(r) & = & \frac{M \ell^2}{3} + (\Xi + \Upsilon) \,, \nonumber \\
x_2(r) & = & \frac{M \ell^2}{3} - \frac{1}{2} (\Xi + \Upsilon) + \mathrm{i} 
\frac{\sqrt{3}}{2} (\Xi - \Upsilon) \,, \label{roots} \\
x_3(r) & = & \frac{M \ell^2}{3} - \frac{1}{2} (\Xi + \Upsilon) - \mathrm{i} 
\frac{\sqrt{3}}{2} (\Xi - \Upsilon) \,. \nonumber
\end{eqnarray}
Defining the critical value $\tilde{r}$ as 
\[ \tilde{r}^2 \equiv \frac{\ell^2}{27 \omega (1-\omega^2)} 
\Big( M \omega \big( 9 + 2 M^2 \omega^2 \big) + 2 \big( 3 + M^2 \omega^2 \big)^{3/2} \Big) > 0 \,, \]
I find that \cite{abrasteg} i) $\Delta>0$ when $r^2 > \tilde{r}^2$, which implies that there exist 
one real and two complex conjugate roots of (\ref{cubic}); ii) $\Delta=0$ when $r = \tilde{r}$, so 
that all roots of (\ref{cubic}) are real and at least two of them are equal; iii) $\Delta<0$
when $0<r^2<\tilde{r}^2$, so that all roots of (\ref{cubic}) are real and unequal. 
Note that since (\ref{cubic}) can be cast as 
\begin{equation} 
x \Big( 1 + \frac{M \, \omega^2}{\ell^2} x - \frac{\omega^2}{\ell^4} x^2 \Big) = (1-\omega^2) r^2 > 0 \,, 
\label{ineq}
\end{equation}
this, though \emph{crudely}, further constrains the metric function $x(r)$ to satisfy either
\begin{equation} 
x(r) <  x_{-} \equiv \frac{M\ell^2}{2} \left( 1 - \sqrt{1+ \frac{4}{M^2 \omega^2}} \, \right) < 0 
\quad \mbox {or} \quad
0 < x(r) < x_{+} \equiv \frac{M\ell^2}{2} \left( 1 + \sqrt{1+ \frac{4}{M^2 \omega^2}} \, \right) \,, 
\label{xpmdef}
\end{equation}
where $x_{\pm}$ denote the roots of the quadratic factor on the left hand side of (\ref{ineq}),
which are different from the \emph{formal roots} (\ref{roots}) of the polynomial (\ref{cubic}). 
The branch \( 0 < x(r) < x_{+} \) allows for an event horizon since \( M \ell^2 < x_{+} \) and 
is the one I use for the interpretation of (\ref{met}) as a stationary Lifshitz black hole. Note 
that this is all the more plausible especially if one takes both $\omega$ and $M$ as small 
but positive, i.e. $\omega \gtrsim 0$ and $M \gtrsim 0$, since then it is easy to see that the 
upper bound of the inequality \( x_{+} \simeq \ell^2/\omega \) can be made arbitrarily large 
with $\ell \gg 1$.

\section{\label{appb}} 
Here I briefly discuss the geodesics of the stationary Lifshitz black hole (\ref{altmetinrho}) 
and compare them to the geodesics of the static Lifshitz black hole (\ref{stamet}). Denoting the
derivatives with respect to an affine parameter with a dot, it immediately follows from 
(\ref{altmetinrho}) that the geodesics satisfy
\begin{eqnarray*}
{\cal L} & = & - \frac{\dot{t}^2}{(1-\omega^2)} 
\left( \frac{\rho^6}{\ell^6} - \frac{M \, \rho^4}{\ell^4} - \frac{\omega^2 \, \rho^2}{\ell^2} \right) 
+ \frac{2 \, \omega \, \ell \, \dot{t} \, \dot{\theta}}{(1-\omega^2)}  
\left( \frac{\rho^6}{\ell^6} - \frac{M \, \rho^4}{\ell^4} - \frac{\rho^2}{\ell^2} \right) \\
& & + \frac{\ell^2 \, \dot{\theta}^2}{(1-\omega^2)} 
\left( \frac{\rho^2}{\ell^2} - \frac{\omega^2 \, \rho^6}{\ell^6} + \frac{M \, \omega^2 \, \rho^4}{\ell^4} \right)
+ \frac{\dot{\rho}^2}{\big( \frac{\rho^2}{\ell^2} -M \big)} \,, 
\end{eqnarray*}
where \( {\cal L} = -1 \) for timelike and \( {\cal L} = 0 \) for null geodesics. For a physical particle
which has energy \( {\cal E} = \partial {\cal L}/\partial \dot{t} = \) const. and orbital angular 
momentum \( {\cal J} = \partial {\cal L}/\partial \dot{\theta} = \) const., the elimination of  $\dot{t}$ 
and $\dot{\theta}$ in terms of ${\cal E}$ and ${\cal J}$ in the obvious way leads to
\begin{eqnarray} 
{\cal L} & = & \frac{1}{4 \rho^2 (1-\omega^2)} \left[
{\cal J}^2 \Big(1+\frac{\ell^4 \omega^2}{\rho^2 (M \ell^2 - \rho^2)} \Big)
+ {\cal E}^2 \ell^2 \Big(\omega^2+\frac{\ell^4}{\rho^2 (M \ell^2 - \rho^2)} \Big) \right. \nonumber  \\
& & \left. \qquad  \qquad \qquad 
+ 2 \ell \omega {\cal E} {\cal J} \Big(1+\frac{\ell^4}{\rho^2 (M \ell^2 - \rho^2)} \Big) \right]
- \frac{\ell^2 \dot{\rho}^2}{(M \ell^2 - \rho^2)} \,. \label{georot}
\end{eqnarray}
In principle from this one can study the solutions of the radial geodesics, but obviously the
resultant expression for $\dot{\rho}$ is quite complicated in the generic case. However, when 
${\cal E}$ and ${\cal J}$ are precisely related by \( {\cal J} + \ell \omega {\cal E} = 0 \), this
simplifies considerably and allows one to arrive at
\begin{equation} 
\dot{\rho}^2 - \big( \frac{\rho^2}{\ell^2} -M \big) {\cal L} = 
\frac{{\cal E}^2 \ell^4 (1-\omega^2)}{4 \rho^4} \,. \label{simprot}
\end{equation}
For a lightlike particle ${\cal L}=0$, and this simplifies further obviously.

If one is to repeat the analogous calculation for the static Lifshitz black hole (\ref{stamet})
and to use \( \bar{\cal E} \) and \( \bar{\cal J} \) for the analogous physical quantities, one 
finds 
\begin{equation}
{\cal L} = \frac{\bar{\cal J}^2}{4 \rho^2} - \frac{\ell^2 \dot{\rho}^2}{(M \ell^2 - \rho^2)}
+ \frac{\bar{{\cal E}}^2 \ell^6}{4 \rho^4 (M \ell^2 - \rho^2)} \,, \label{geosta}
\end{equation}
which unsurprisingly amounts to setting $\omega = 0$ and barring the relevant bits in (\ref{georot}).
In this case when the orbital angular momentum \( \bar{\cal J} = 0 \), the analog of (\ref{simprot})
becomes
\begin{equation} 
\dot{\rho}^2 - \big( \frac{\rho^2}{\ell^2} -M \big) {\cal L} = 
\frac{\bar{{\cal E}}^2 \ell^4}{4 \rho^4} \,, \label{simpsta}
\end{equation}
which again amounts to setting $\omega = 0$ and barring ${\cal E}$ in (\ref{simprot}). Once again,
for lightlike particles this becomes very simple.

By now, the upshot of all this discussion should be clear. The rotation introduced by $\omega$
indeed complicates the behavior of the geodesics, but not by a great margin. One can
easily conclude, at least for the simple geodesics discussed above, that whatever attributes 
the geodesics of the static Lifshitz black hole (\ref{stamet}) possess the same attributes are, 
more or less, also shared by the stationary Lifshitz black hole (\ref{altmetinrho}). In particular,
one may be tempted to conclude that a timelike particle with a large enough ${\cal E}$ can go
from the horizon $\rho^2 = \ell^2 M$ to the timelike surface $\rho_{1} \equiv x_{1}^2$ 
(\ref{xpmdef}) within a finite affine parameter interval and proceed further to larger values 
of $\rho = x^2$ before being reflected back at some turning point  \( x_{{\cal E}} > x_{+} \), but
such a thing is impossible as per what happens for the analogous case of the static Lifshitz black 
hole, and all the more so for the choices $\omega \gtrsim 0$, $M \gtrsim 0$ and $\ell \gg 1$.

\end{document}